\documentclass[preprint,11pt]{revtex4-1}
\usepackage[utf8]{inputenc}
\usepackage[T1]{fontenc}
\usepackage{bm}
\usepackage{amsmath,ulem}
\usepackage{amsfonts}
\usepackage{amssymb}
\usepackage{graphicx}
\usepackage{siunitx} 
\usepackage{float}
\makeatletter
\let\saved@includegraphics\includegraphics
\AtBeginDocument{\let\includegraphics\saved@includegraphics}
\makeatother
\RequirePackage{siunitx}
\usepackage{xcolor}

\newcommand{\mum}{\si{\micro\metre}}
\begin{document}
\title{Harnessing Synthetic Active Particles for Physical Reservoir Computing}
\author{Xiangzun Wang$^{1,2}$ and Frank Cichos${}^{1\ast}$\\
\textit{${}^{1}$Peter Debye Institute for Soft Matter Physics, Leipzig University, 04103 Leipzig, Germany}\\
\textit{${}^{2}$Center for Scalable Data Analytics and Artificial Intelligence (ScaDS.AI) Dresden/Leipzig, 04105 Leipzig, Germany}\\
\textit{$^*$To whom correspondence should be addressed; E-mail: cichos@physik.uni-leipzig.de}
}

\begin{abstract}
The processing of information is an indispensable property of living systems realized by networks of active processes with enormous complexity. They have inspired many variants of modern machine learning one of them being reservoir computing, in which stimulating a network of nodes with fading memory enables computations and complex predictions. Reservoirs are implemented on computer hardware, but also on unconventional physical substrates such as mechanical oscillators, spins, or bacteria often summarized as physical reservoir computing.
Here we demonstrate physical reservoir computing with a synthetic active microparticle system that self-organizes from an active and passive component into inherently noisy nonlinear dynamical units. The self-organization and dynamical response of the unit is the result of a delayed propulsion of the microswimmer to a passive target. A reservoir of such units with a self-coupling via the delayed response can perform predictive tasks despite the strong noise resulting from Brownian motion of the microswimmers. To achieve efficient noise suppression, we introduce a special architecture that uses historical reservoir states for output. Our results pave the way for the study of information processing in synthetic self-organized active particle systems.
\end{abstract}


\maketitle

\section*{Introduction}
Storing and processing of information is vital for living systems~\cite{Tkacik.2014}. The detection of low amounts of chemicals by a  bacterium to navigate environments~\cite{Wadhams2004,Celani.2010}, the feedback mechanisms controlling and maintaining the function of organisms~\cite{Cosentino.2019}, or the highly sophisticated computations in large biological neural networks in the brain~\cite{Knudsen.1987} are intricate examples of this importance and created by evolutionary development. All these processes with living matter as the substrate of computation rely on its inherent activity, e.g., the microscopic energy conversion to power the signalling cascades in the presence strong thermal noise. They have inspired many computational models of machine learning that are not executed on living matter, but on well-designed electronic hardware using completely different information representation than living matter~\cite{Marković.2020}. Recurrent neural networks are a variant of such mathematical algorithms with a fading memory that allow learning from information sequences as in language or time series~\cite{Abiodun2018}. Reservoir computers employ sparsely and statically connected recurrent nodes~\cite{Jaeger2001,Maass2002,Verstraeten2007} or even a single node by using time-multiplexing~\cite{Paquot.2010,Appeltant2011} to create a high dimensional space. Information can be injected into this space to spread over the many degrees of freedom. Unlike the training of other neural networks, where the interactions of all components are optimized, training reservoir computers is often only restricted to finding how the desired information can be retrieved from the node states using adjustable readout only~\cite{Lukosevicius2009,Lukosevicius2012}. 

As one of the main properties of recurrent nodes is the memory of past states, reservoir computers also allow for a physical realization on unconventional computational substrates~\cite{Tanaka2019,Nakajima2020,Rafayelyan.2020} using optoelectronic oscillators~\cite{Larger2012,Paquot2012,VanDerSande2017}, mechanical oscillators~\cite{Coulombe2017,Dion2018}, carbon nanotubes~\cite{Wu.2021} or passive soft bodies~\cite{Tanaka2019} as excitable physical systems. It is thus intruiging to close the loop and draw inspiration from active living systems to explore microscopic reservoir computing in synthetic active microsystems, where noise is omnipresent as well but a precise control over the shape and the physics of the active system is possible. Motile synthetic active particles have generated enormous interest as a model for self-propelled systems far from equilbrium and emergent collective effects~\cite{Bechinger.2016fj9}, that mimic, for example, the dynamics of swarms~\cite{Lavergne.2019,Baeuerle.2020,Wang2023}. Information processing~\cite{Woodhouse.2017,Colen.2021} and learning~\cite{Cichos2020} in experiment~\cite{Muinos-Landin2021} and simulation~\cite{Liebchen.2019,Schneider.2019,Colabrese.201760u,Colabrese.2018,Gustavsson.2017} have also entered the field of synthetic active matter. However, studies that extend the use of synthetic microparticles as computational substrates are still rare or purely computational~\cite{Lymburn2021}. 

We demonstrate that motile self-propelled active microparticles can be used for physical reservoir computing. An active particle self-organizes into a nonlinear dynamical unit based on a retarded propulsion towards an immobile target forming a noisy physical recurrent node. The node is perturbed by a time-multiplexed input signal to form a network of virtual nodes with sparsely connected topology. Multiple of these active units realize a high dimensional space of our reservoir computer. Harnessing the physics and inherent dynamics of the active particles, the reservoir computer is capable of predicting chaotic signals despite the strong influence of Brownian motion of the active particles. In particular, we find, that using historical reservoir states for the output derivation effectively suppresses the intrinsic noise of the reservoir opening new routes for reservoir computing in noisy systems.

\section*{Results}
\subsection*{Active particle recurrent node} \label{sec:principle}
\begin{figure}[ht]
    \centering
    \includegraphics[width=0.98\columnwidth]{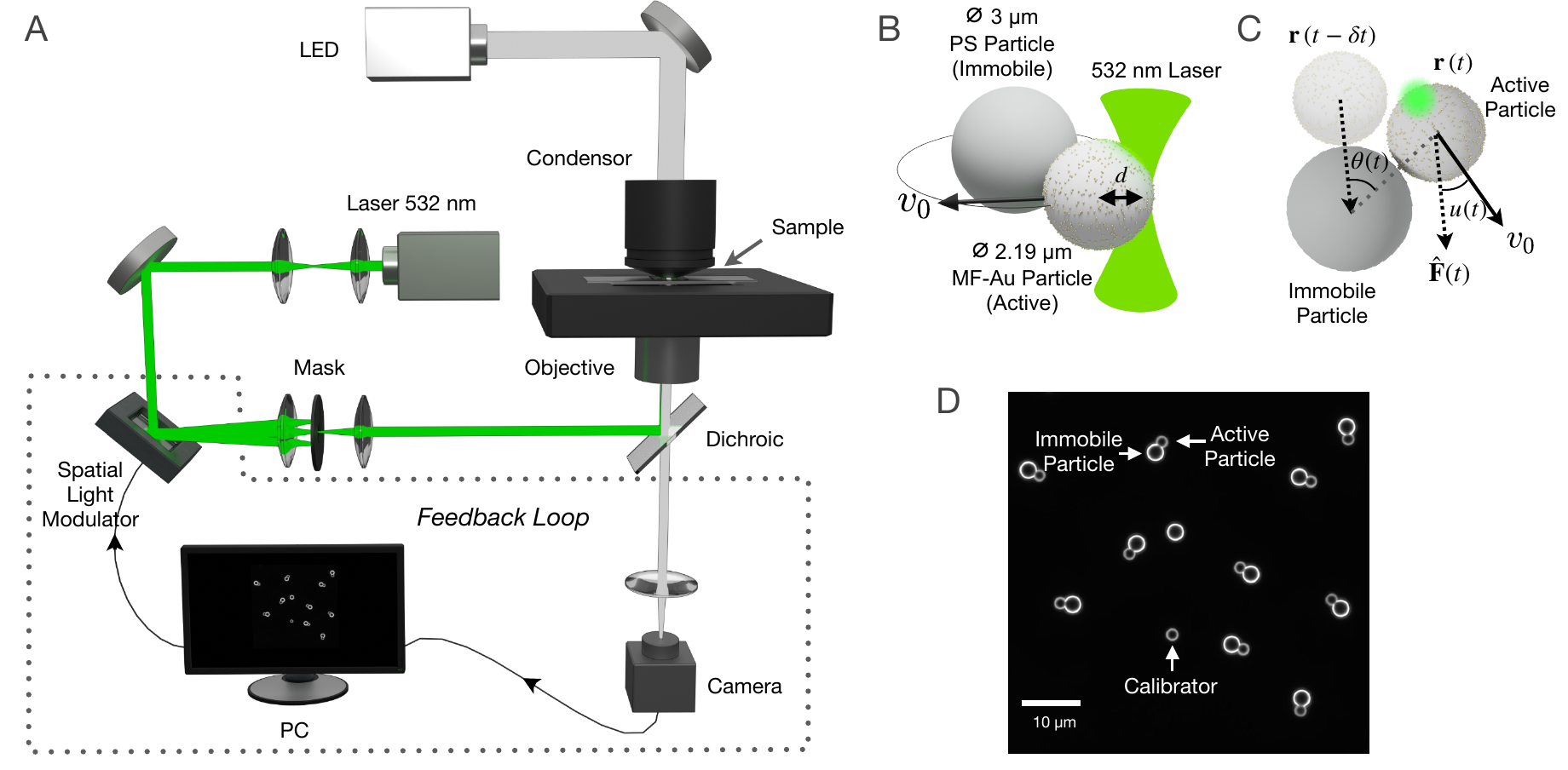}
    \caption{
\textbf{Experimental realization.}
    \textbf{A} Experimental setup (see Sec.~Method
    for detail). 
    \textbf{B} Active particle recurrent unit consisting of a gold-nanoparticle covered melamine resin particle (MF-Au) and an immobile target particle (PS). For active particle propulsion a $532~\si{\nano\metre}$ laser is focused to a distance d $d$ from the particle center. The resulting heat and asymmetric temperature induce a self-thermophoretic motion of the particle with a speed of $v_0$ and a direction set the vector from the laser to the particle center.
    \textbf{C} Top view of the active particle system. The active particle is controlled to carry out a motion towards the immobile particle along $\hat{\mathbf{F}}(t)$ with a time delay $\delta t$ in its reaction. The direction of $\hat{\mathbf{F}}(t)$ (dashed arrow) is determined by the previous active particle location $\mathbf{r}(t-\delta t)$. An additional angle $u(t)$ between the particle propulsion direction (solid arrow) and $\hat{\mathbf{F}}(t)$ represents an external input into the system. 
    \textbf{D} Darkfield microscopy image of the sample consisting of 10 active-immobile particle pairs (larger circle is the immobile particle, smaller circle the active particle) as physical nodes in the experiment. An additional calibrator is used as an active particle swimming along a square route to measure the propulsion speed $v_0$. The real-time video is provided in Supplementary Movie 1. 
    }\label{fig:1}
\end{figure}

A reservoir computer (RC), as a paradigm derived from recurrent neural networks, consists of recurrent nodes that nonlinearly process external signal inputs as well as their previous outputs~\cite{Jaeger2001,Verstraeten2007}. We realize a simple recurrent node with the help of a single synthetic active particle as a microscopic model for motile active matter~\cite{Bechinger.2016fj9}. The active particle is a polymer microbead of $2.19\,\si{\micro\metre}$ diameter with the surface decorated by gold nanoparticles (about $8\,\si{\nano\metre}$ diameter). It is immersed in a thin layer of water bounded by two cover glasses and can move freely in two dimensions. It is propelled with a speed $v_0$~\cite{Franzl2021} towards an immobile target by partial heating of the gold nanoparticles using a focused laser in a microscopy setup (see Fig.~\ref{fig:1}A,~B and Methods section). The continuous propulsion in a desired direction is realized by adjusting the focused laser spot position in real time with the help of a feedback loop using a spatial light modulator (SLM). A time delay $\delta t$ is added to the intrinsic feedback latency to control the active particle~\cite{Khadka2018,Wang2023} and to introduce a finite reaction time, as is inherent in many processes in living systems. Correspondingly, the attraction $\hat{\mathbf{F}}(t)$ experienced by the particle at time $t$ is determined by its position at previous time $t-\delta t$, 
\begin{equation} 
    \hat{\mathbf{F}}(t)  = -\frac{\mathbf{r}(t-\delta t)}{|\mathbf{r}(t-\delta t)|},
    \label{eq:delay force}
\end{equation}
where $\mathbf{r}$ denotes the location of the active particle with respect to the immobile particle center. The consequence of this retardation is a particle angular displacement $\theta(t) = \phi(t) - \phi (t-\delta t)$ during the delay time, which results in a transient rotational motion of the active particle around the immobile target (Fig.~\ref{fig:1}C), where $\phi$ denotes the angular position of the particle. The dynamics of the angular position can then be described by a nonlinear delay differential equation 
\begin{equation}
    \dot{\phi}(t) = \frac{v_0}{R_0}\sin \big( \theta(t) + u(t) \big) + \frac{\sqrt{2D}}{R_0}w(t)
    \label{eq:phi dynamics}
\end{equation}
assuming that the active and the immobile particle (radius of $a_\mathrm{act}$ and $a_\mathrm{imm}$) are in physical contact, i.e. $R_0 = a_\mathrm{act} + a_\mathrm{imm}$, due to the delayed attraction. The active particle in water is subject to Brownian motion as represented by the noise term in Eq.~\ref{eq:phi dynamics} with $w(t)$ denoting Gaussian white noise. The diffusion coefficient was determined from the experiment to be $D$ = 0.08~µm$^2$s$^{-1}$, giving rise to a Pećlet number of $\mathrm{Pe}$ = $a_\mathrm{act}v_0/D$ = 38.7. To make the physical recurrent node capable of receiving external inputs, we introduce the $u(t)$ in Eq.~\ref{eq:phi dynamics} representing an angular deviation of the particle propulsion direction from $\hat{\mathbf{F}}$ (Fig.~\ref{fig:1}C). 
For its function as a recurrent node, the dynamics of the angle $\theta(t)$ is important. The dynamics described by Eq. \ref{eq:phi dynamics} can be approximated by an overdamped motion of a particle in a self-generated effective quartic potential $U_{\rm eff}(\theta)$~\cite{Wang2023}, which resembles a generic Landau-type description~\cite{Goldenfeld.2018,Wang2023}. The shape of the potential is controlled by a dimensionless parameter $\theta_0=v_0\delta t/R_0$. With increasing $\theta_0$, $U_{\rm eff}(\theta)$ transitions from a near parabolic shape with a minimum at $\theta=0$ to a symmetric double well shape with minima at $\theta_+$ and $\theta_-$ in a pitchfork bifurcation (see Fig.~\ref{fig:2}C and Sec.~S2 of Supplementary Information).

A perturbation of $\theta$ (solid arrow in Fig.~\ref{fig:2}C), will result in a relaxation with a dynamics determined by the control parameter $\theta_0$. We have experimentally determined the response $\theta(t)$ to an impulsive perturbation $u(t)=\delta(t)$ for different values of $\theta_0$ (Fig.~\ref{fig:2}A,~B).
The individual experimental trajectories $\theta(t)$ (grey lines in Fig.~\ref{fig:2}A) strongly fluctuate due to the Brownian motion of the active particle. However, the ensemble average of 500 trajectories of $\theta(t)$ (red lines in Fig.~\ref{fig:2}A and solid lines in B) exhibits an asymptotic behavior, which nicely reflects the response evaluated in deterministic simulations (Fig.~\ref{fig:2}B, dashed lines). The characteristic relaxation time $\tau_{\theta}$ extracted from the deterministic simulations reveals the expected strong increase around the transition point. Tuning the control parameter $\theta_0$, i.e. activity ($v_0$) and/or delay ($\delta t$) therefore allows to manipulate the fading memory of the active particle recurrent node, which is paramount for the coupling and response of the recurrent nodes in our reservoir computer. Note, that the nonlinear dynamics of the active particle system is the result of the delayed propulsion towards the target.

\subsection*{Reservoir computer with active particle nodes}

\begin{figure}[h]
    \centering
    \includegraphics[width=0.85\columnwidth]{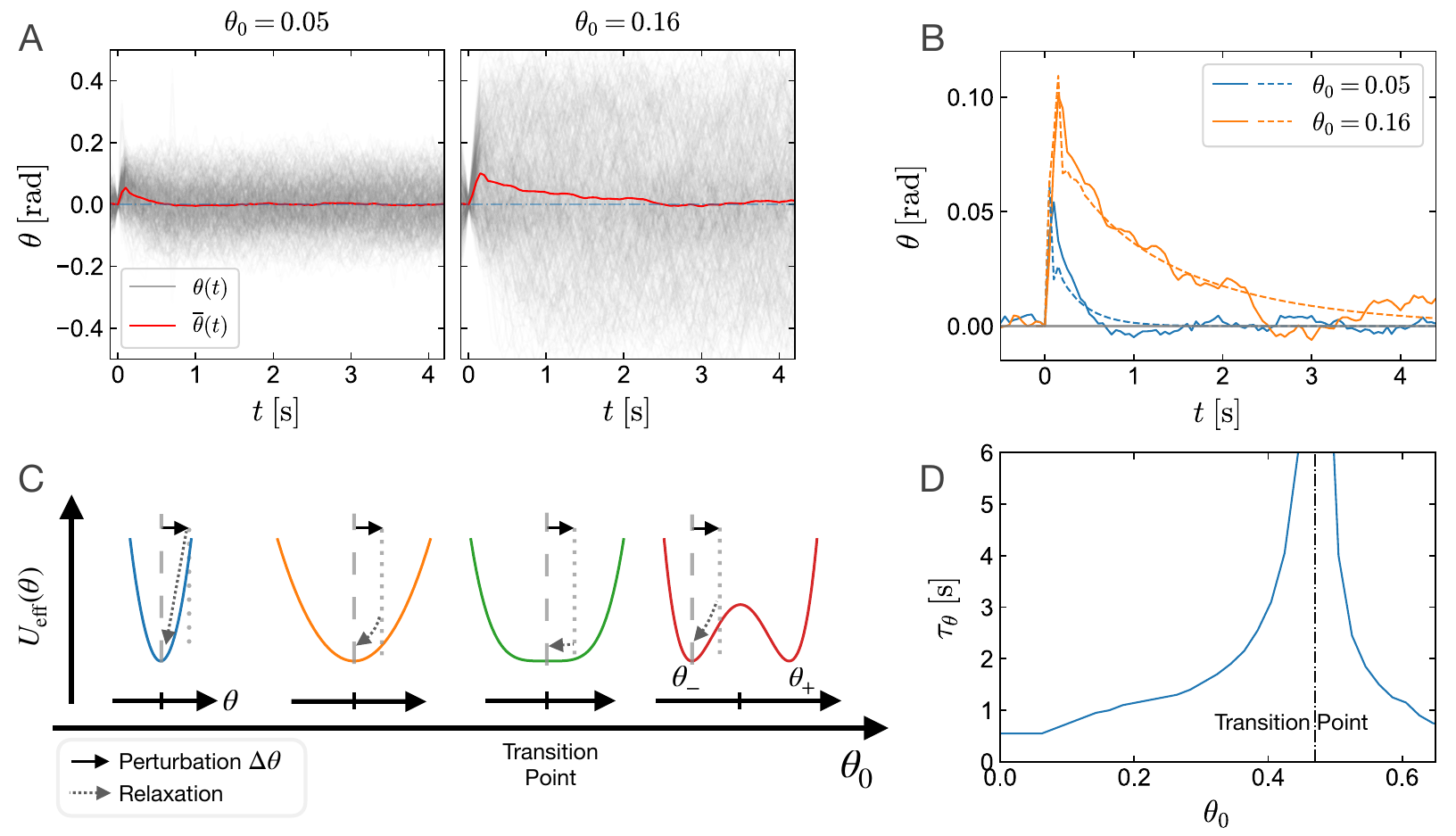}
    \caption{
\textbf{Impulse response of teh active particle node.}
        \textbf{A} Respons of $\theta(t)$ to an impulsive input $u(t)$ = $\delta(t)$ measured in the experiment with $v_0\delta t/R_0$ of 0.05 and 0.16. The grey curves denote the measured $\theta(t)$ traces. They strongly fluctuate due to  the Brownian motion of the active particles. The red curves denote the means of 500 $\theta(t)$ traces in each case. \textbf{B} Comparison of the mean impulse responses of $\theta(t)$ from the experiment (solid lines, same as in panel (\textbf{A})) to the ones obtained from the deterministic simulation (dashed lines). \textbf{C} Effective potential $U_\mathrm{eff}(\theta)$ with different $v_0\delta t/R_0$. $U_\mathrm{eff}(\theta)$ transitions from a single well to a double well form at the transition point, with its local minima positions bifurcating from zero to two opposite values ($\theta_{+}$, $\theta_-$). After perturbations $\Delta \theta$ (solid arrows), $\theta$ relaxes to one of the minima (dashed arrows). 
        \textbf{D} Relaxation time $\tau_\theta$ of $\theta$ as function of $v_0\delta t/R_0$ evaluated in deterministic simulations with $\delta t = 0.6\,\si{\second}$ varying $v_0$ from $0\,\si{\micro\metre\per\second}$ to $2.74\,\si{\micro\metre\per\second}$. $\theta$ is perturbed with $\Delta \theta$ from one of the states $\theta_{+,-}$ at $t = 0\, \si{\second}$, then relaxes to $\theta(\tau_\theta)-\theta_{+,-}$ = $\Delta\theta/10$ . $\tau_\theta$ diverges to infinity at the transition point of $U_{\rm eff}(\theta)$ (dash-dotted line) at $v_0\delta t/R_0 = 0.46$.
    }\label{fig:2}
\end{figure}

\begin{figure}[ht]
    \centering
    \includegraphics[width=0.95\columnwidth]{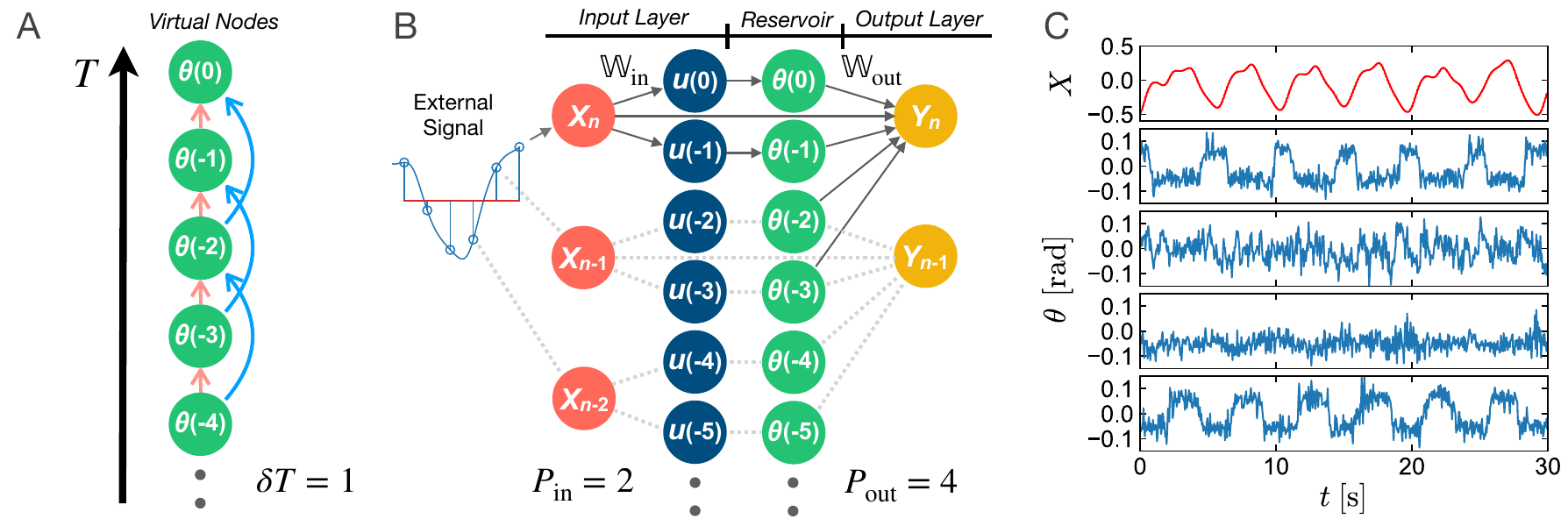}
    \caption{
    \textbf{Architecture of the reservoir computer.}
    \textbf{A} Topology of the reservoir with a single physical node with a discrete-time delay $\delta T = 1$ in this example. The angle $\theta$ of the active particle at different timesteps $T$ is considered as a virtual node. 
    Each virtual node state $\theta(T)$ is nonlinearly coupled to previous states $\theta(T-1)$ and $\theta(T-\delta T-1)$ through the physics of the system (Eq.~\ref{eq:theta evolution}).  
    \textbf{B} Sketch of the information processing in the single node RC with $P_\mathrm{in} = 2$ and $P_\mathrm{out} = 4$ in this example. The external signal $X$ of each computation step is multiplexed by a weight matrix $\mathbb{W}_\mathrm{in}$ to $P_\mathrm{in}$ elements, which are sequentially input into the node as the perturbation $u(T)$ of the active particle (Fig.~\ref{fig:1}C). Each output is linearly derived from the $\theta$ states of previous $P_\mathrm{out}$ timesteps using a weight matrix $\mathbb{W}_\mathrm{out}$ that is trained using ridge regression. Biases $b_\mathrm{in/out}$ for the input and output layers are not plotted for the sake of simplicity. 
    \textbf{C} Examples of $\theta(t)$ traces of four physical nodes (blue lines) driven by an external signal $X(t)$ (red line, see Eq.~\ref{eq:Mackey Glass}) measured in the experiment.}
    \label{fig:3}
\end{figure}

The asymptotic relaxation of $\theta(t)$ demonstrated in Fig.~\ref{fig:2}B represents a basic requirement for reservoir computing~\cite{Boyd1985,Maass2002,Maass2004,Dambre2012}. Based on the time-delay it also allows the setup of coupled recurrent nodes. In the discrete-time setting of our experiment with the sampling period $\Delta t$, Eq.~\ref{eq:phi dynamics} can be rewritten as
\begin{equation}
    \phi(T)  = \phi(T-1) + \frac{v_0\Delta t}{R_0}\sin \big( \theta(T-1) + u(T-1) \big) + \frac{\sqrt{2D\Delta t}}{R_0} W(T-1),
    \label{eq:dphi_dr_discrete}
\end{equation}
where $T$ = $t/\Delta t$ is an integer number representing the timestep. $W$ denotes Gaussian random numbers with zero mean and unit variance. The evolution of $\theta$ then follows as
\begin{equation}
    \begin{aligned}
        \theta(T) = \theta(T-1) &+ \frac{v_0\Delta t}{R_0}\sin\big(\theta(T-1) + u(T-1)\big) \\
        	&- \frac{v_0\Delta t}{R_0}\sin\big(\theta(T-\delta T-1) + u(T-\delta T-1)\big) \\
         &+ \frac{\sqrt{2D\Delta t}}{R_0} \big(W(T-1) - W(T-\delta T-1) \big),
    \end{aligned}
    \label{eq:theta evolution}
\end{equation}
with the discrete-time delay $\delta T$ = $\delta t/\Delta t$. Refering to the concept of virtual nodes and time-multiplexing~\cite{Appeltant2011}, we consider the transient state $\theta(T)$ of a physical node at different timesteps as virtual nodes constituting the reservoir. Each virtual node state $\theta(T)$ is, according to Eq.~\ref{eq:theta evolution}, coupled to its previous states $\theta(T-1)$ and $\theta(T-\delta T-1)$. The virtual nodes thus reflect a topology with sparse interconnections realized by the delay of the physical node (Fig.~\ref{fig:3}A). The interconnections are inherently nonlinear due to the \textit{sine} function in Eq.~\ref{eq:theta evolution}, which originates from the physical interaction between the active and the immobile particle and naturally serves as the activation function in our RC. 

The working principle of our RC is now illustrated in Fig. \ref{fig:3}B. For simplicity of the discussion, we consider a RC with a single physical node, scalar input $X_n$ and output $Y_n$ at the $n$-th computation step (for general case see Sec.~4.1 of Supplementary Information). The input layer $\mathbf{u}_n$ is generated via a matrix of input weights $\mathbb{W}_\mathrm{in}\in\mathbb{R}^{P_\mathrm{in}\times 2}$ 
\begin{equation}
    \mathbf{u}_n = \mathbb{W}_\mathrm{in}[b_\mathrm{in},\,X_n]^\mathrm{T},
\end{equation}
with the scalar $b_\mathrm{in}$ as the input bias. $\mathbf{u}_n$ is a one-dimensional array containing $P_\mathrm{in}$ elements, which are sequentially input into the physical node as the angle $u(T)$ of the active particle in Eq.~\ref{eq:dphi_dr_discrete}. This operation is equivalent to a time-multiplexing of the input $X_n$ into $P_\mathrm{in}$ virtual nodes with $\mathbb{W}_\mathrm{in}$ as the mask, as commonly applied in continuous-time single node physical RC approaches~\cite{Appeltant2011,Appeltant2014}. The output $Y_n$ is derived as a linear combination of a scalar bias $b_\mathrm{out}$, the input signal $X_n$, and the node states $\theta$ of the past $P_\mathrm{out}$ timesteps using an output weight matrix $\mathbb{W}_\mathrm{out}\in\mathbb{R}^{1\times(2+ P_\mathrm{out})}$
\begin{equation}
    Y_n = \mathbb{W}_\mathrm{out}[b_\mathrm{out},\,X_n,\,\theta(T=0), \,\theta(T=-1),\cdots,\, \theta(T=1-P_\mathrm{out})]^\mathrm{T},
\end{equation}
where $T = 0$ denotes the time of the current computation step. The output weight matrix $\mathbb{W}_\mathrm{out}$ is the only quantity to be trained in the RC framework to tune the output towards the target signal. It is calculated via a ridge regressions~\cite{Lukosevicius2012} in each computation cycle.

As compared to conventional time-multiplexed physical RCs where $P_\mathrm{in}$ = $P_\mathrm{out}$, we explicitly allow these two parameters to be independent and freely adjustable (see Fig.~\ref{fig:3}B). In particular, we set $P_\mathrm{out} \gg P_\mathrm{in}$ for our RC. By this means, each output is not only derived from the reservoir states of its corresponding step, but also from previous $n_\mathrm{hist}$ = $P_\mathrm{out}/P_\mathrm{in}-1$ steps. It will be demonstrated in the next sections that this setting enables us to carry out the RC with a good stability of the output and, most importantly, an effective reduction of the impact of the intrinsic noise. 
This single physical node architecture can be further extended to multiple physical nodes operated in parallel. In our experiment, we control $N_\mathrm{node} = 10$ independent physical nodes in one sample simultaneously. The transient state $\theta$ of the 10 physical nodes together constitutes the virtual nodes of the reservoir. Fig.~\ref{fig:1}D and Supplementary Movie 1 show the real time image and video of the sample in the experiment. Fig.~\ref{fig:3}C plots an exemplary trace of $\theta(t)$ of four physical nodes (blue lines) driven by an external signal $X$ (red line) in the experiment. 

The configuration of the RC is optimized in a simulation for the best performance and then applied to the experiment. The input weights are selected from a binary distribution $\mathbb{W}^i_\mathrm{in}$ $\in\{-2,2\}$, which results in a better noise resistance of the RC than using $\mathbb{W}^i_\mathrm{in}$ $\in\{-1,1\}$. The details of the RC configuration are described in Sec.~4 of Supplementary Information.

\subsection*{Chaotic series prediction} \label{sec:result}

We test our RC with the free-running prediction of the chaotic Mackey-Glass series (MGS). The MGS is generated by the delay differential equation 
\begin{equation}
    \frac{\,\mathrm{d} S(n)}{\,\mathrm{d} n} = \alpha \frac{S(n-\tau)}{1+S(n-\tau)^{\beta}} - \gamma S(n),
    \label{eq:Mackey Glass}
\end{equation}
which was introduced to model the complex dynamics of physiological feedback systems~\cite{Mackey1977}. It has been widely used as a benchmark task for series forecasting~\cite{Jaeger2004,Antonik2017b,Fujii2017,Dong2020}. With parameters $\alpha$ = 0.2, $\beta$ = 10, $\gamma$ = 0.1, and a delay parameter $\tau$ = 17, the MGS exhibits a chaotic behavior with the a Lyapunov exponent of around 0.006~\cite{Jaeger2004}. The performance of the prediction is evaluated by the normalized root-mean-square error (NRMSE, see Sec.~S4, S5 of Supplementary Information for details). Fig.~\ref{fig:4} shows the results of the MGS predictions by our RC in experiments and simulations. 

\paragraph*{\bf Simulated prediction}
The deterministic simulations have been carried out with a very small reservoir with only $N_\mathrm{node}P_\mathrm{in}$ = 20 virtual nodes but using a large $P_\mathrm{out}$ = 400, i.e., $n_\mathrm{hist}$ = 199 historical reservoir states for each output.
The simulation result shows a very good prediction of the target MGS up to around 900 steps (corresponding to 5.4 Lyapunov time) with a NRMSE of 6.7$\times$10$^{-2}$ (Fig.~\ref{fig:4}A). Similarly, Sec.~S7.2 of the Supplementary Information also describes the prediction of a three-dimensional chaotic Lorenz series by our RC in a deterministic simulation. These results signify the capability of our architecture for chaotic systems predictions.

\paragraph*{\bf Experimental prediction}
As compared to the simulations, the experimental RC performance is significantly degraded as a result of the Brownian motion of the active particles and the sensitivity of chaotic systems. The noise acts on the particle angular position $\phi(T)$ (Eq.~\ref{eq:dphi_dr_discrete}) and propagates to the virtual node states $\theta(T)$ (Eq.~\ref{eq:theta evolution}). The signal-to-noise ratio (SNR) of the RC can estimated by comparing the angular displacement of the active particle between neighboring timesteps $\Delta \phi(T) = \phi(T) - \phi(T-1)$ in the simulation with and without noise (see Sec.~S6 of Supplementary Information) and reveals an extremely low  value (SNR = 1.9 (2.78 dB)). Fig.~\ref{fig:4}B depicts the experimental results of 50 MGS predictions with identical configurations ($\mathbb{W}_\mathrm{out}$ trained for each repetition). The outputs $Y(n)$ (gray curves) from different experimental predictions exhibit a notable influence of the noise. The mean of the outputs $\overline{Y}(n)$ (blue curve) can reproduce the fundamental period of the target MGS for round 200 steps and details of the series for about 50 steps, which is, considering the extremely low SNR, still remarkable. These results obtained for the experimental RC are further underscored by the very good prediction of a non-chaotic periodic trigonometric series (see Sec.~S7.1 of Supplementary Information). 
\begin{figure}[!h]
    \centering
    \includegraphics[width=0.95\columnwidth]{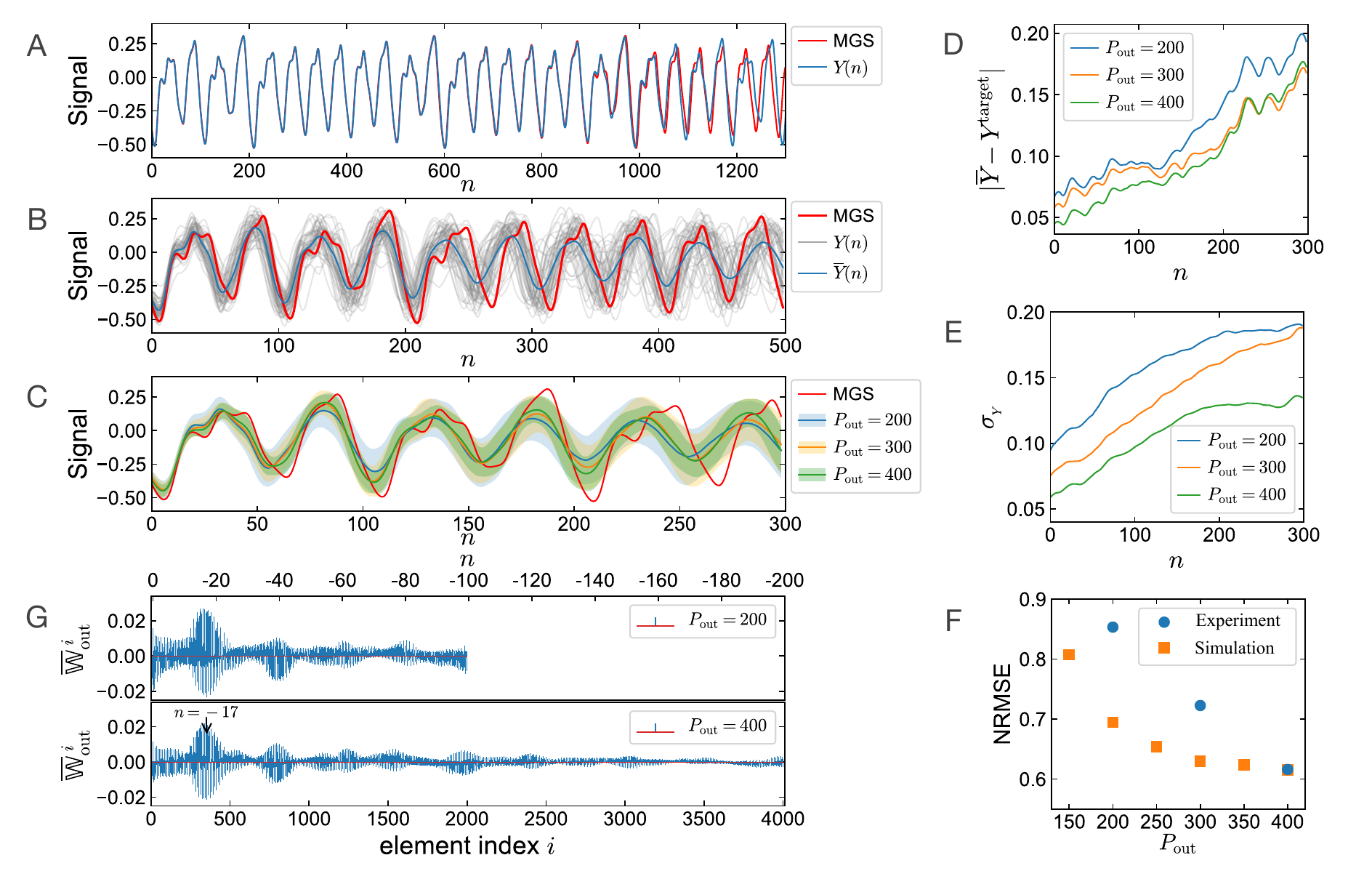}
    \caption{
    \textbf{Results of free-running predictions of the Mackey-Glass series.} 
    \textbf{A} Prediction (blue line) of the target Mackey-Glass series (MGS, red line) by the RC in the deterministic simulation using $N_\mathrm{node}P_\mathrm{in}$ = 20 virtual nodes and $n_\mathrm{hist}$ = 199 historical states (i.e. $P_\mathrm{out}$ = 400). The simulations use a active particle speed of $v_0=2.78\,\si{\micro\metre\per\second}$ and a delay of $\delta t = 0.05\,\si{\second}$. Other parameters of the RC are given in Sec.~4.3 of Supplementary Information. 
     \textbf{B} Experimental results of MGS predictions with the same RC configuration as in panel (\textbf{A}). The gray curves denote the RC outputs from 50 repeated predictions with strong fluctuations due to Brownian motion. 
     The blue curve 
     represents the mean of the output traces. 
    \textbf{C--G} Comparison of the experimental results with $P_\mathrm{out}$ from 200 to 400 evaluated by 50 repeated predictions for each $P_\mathrm{out}$. 
    \textbf{C} Means (colored lines) and corresponding standard deviations (colored areas) of the RC output traces. 
    \textbf{D} The deviation between the mean of predictions and the target MGS, and 
    \textbf{E} the standard deviations $\sigma$ of the predictions versus the step $n$. The curves are smoothed via 100-step moving average. 
    \textbf{F} NRMSE of 200 steps predictions from experiments and stochastic simulations versus $P_\mathrm{out}$. 
    \textbf{G} Means of the output weights $\mathbb{W}_\mathrm{out}$ trained in experiments. 
    The elements of $\mathbb{W}_\mathrm{out}$ are in turn the weight for the output bias $b_\mathrm{out}$, the input signal $X$, and the historical $\theta$ states of the physical nodes (see Sec.~S4.1 of Supplementary Information). Each $N_\mathrm{node}P_\mathrm{in}$ = 20 weights for $\theta$ correspond to one computation step, which is denoted by $n$ on the top axis 
    with the negative values representing the past. The peak marked by the arrow indicates the high contribution of the historical reservoir states of around $n$ = -17, which reveals the property of the target MGS with the delay parameter $\tau$ = 17 (Eq.~\ref{eq:Mackey Glass}).
    }\label{fig:4}
\end{figure}
The performance under these very noisy conditions becomes possible by the special architecture using a number of historical states $n_\mathrm{hist}$ for the output derivation. Fig.~\ref{fig:4}C--G demonstrate the experimental results of the RC with $P_\mathrm{out}$ varying from 200 to 400, corresponding to $n_\mathrm{hist}=99\ldots 199$ historical states. The accuracy of the mean of the predictions is improved by increasing $P_\mathrm{out}$ (Fig.~\ref{fig:4}C, colored lines). 
Changing $P_\mathrm{out}$ from 200 to 400 results in a decrease of the difference between $\overline{Y}(n)$ and the target MGS by 22\% (evaluated by the root-mean-square (RMS) of 200 steps $|\overline{Y}$ - $Y^\mathrm{target}|$ in Fig.~\ref{fig:4}D). The standard deviation $\sigma_{Y}$ of the prediction decreases by 32\% (200 steps RMS of $\sigma_{Y}$ in Fig.~\ref{fig:4}E) and the resulting NRMSE of 200 step predictions diminishes by 28\%. This trend of lower prediction error for the experimental system is also picked up by stochastic simulations with an NRMSE improvement of 11\% (Fig.~\ref{fig:4}F). 

This finding is striking as increasing $P_\mathrm{out}$ does not add more information to the RC, nor increases the reservoir dimensionality, which is determined by the number of independent variables~\cite{Dambre2012} of the reservoir. During the computation of the RC, the node states of each step are nonlinearly transformed and mapped into the states of the following steps~\cite{Dambre2012}, while they attenuate in magnitude due to the fading memory. Conventional RCs derive the output from the reservoir state of its corresponding step, which implicitly contains the information of the historical states. Whereas in our RC, the historical reservoir states can directly contribute to the output. 

The actual contributions of the historical states are determined via the training of the output weights $\mathbb{W}_\mathrm{out}$. The states correlating more to the current output obtain higher weight magnitudes. Fig.~\ref{fig:4}G plots the $\mathbb{W}_\mathrm{out}$ trained in experiments with $P_\mathrm{out}$ of 200 and 400. The first 2000 elements of $\mathbb{W}_\mathrm{out}$ corresponding to the near past reservoir states ($n$ from 0 to -99) have similar structures in both cases. As compared to the near past, the far past states ($n$ from -100 to -199 for $P_\mathrm{out}$ = 400) correspond to smaller weights in amplitude, indicating the weaker correlations to the current output. The highest magnitude of $\mathbb{W}_\mathrm{out}$ appears at around $n$ = $-$17, which coincides with the delay parameter $\tau$ = 17 of the target MGS (Eq.~\ref{eq:Mackey Glass}). The trained $\mathbb{W}_\mathrm{out}$ thereby partly reveal the property of the target signal.

\begin{figure}[!h]
    \centering
    \includegraphics[width=0.95\columnwidth]{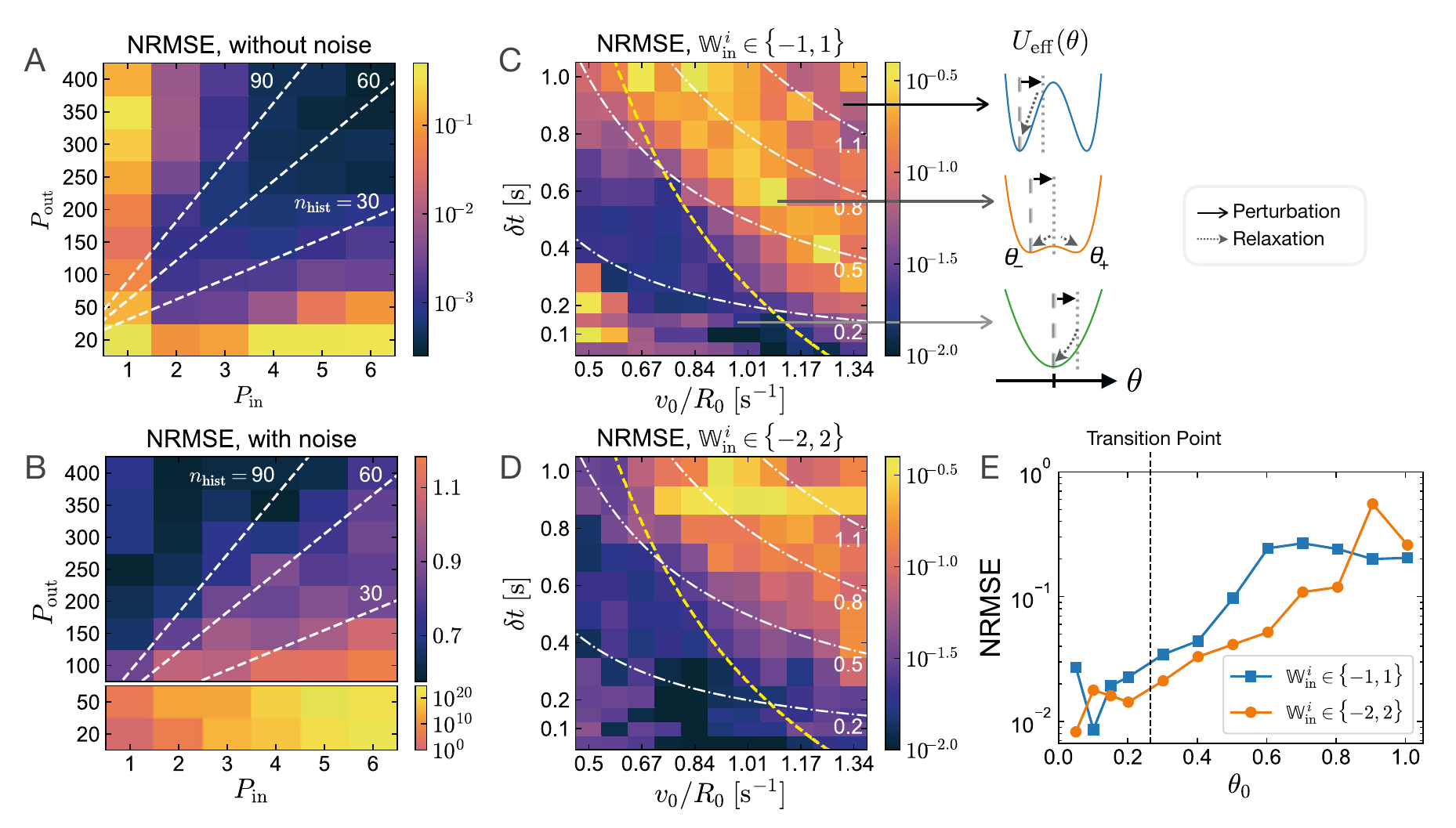}
    \caption{
    \textbf{Simulation results of the RC performance as function of the RC configuration.} 
    NRMSE of 200 steps MGS prediction versus $P_\mathrm{in}$ and $P_\mathrm{out}$ (\textbf{A--B}), $v_0/R_0$ and $\delta t$ (\textbf{C--E}). $\mathbb{W}_\mathrm{in}$ and $b_\mathrm{in}$ are optimized (see Sec.~S4.3 of supplementary Information) 
    for each grid point. 
    \textbf{A} Results of RC without noise. The white dashed lines represent the contour lines of $n_\mathrm{hist} = P_\mathrm{out}/P_\mathrm{in}-1$. 
    \textbf{B} Results of RC with noises and the same parameters as in panel (\textbf{A}). Each grid point is evaluated by 50 repeated predictions. The outputs of RC with $P_\mathrm{out}<100$  
   are unstable, and result in large NRMSE as plotted separately in the bottom subplot. 
    \textbf{C,~D} Results of RC without noise. The elements of $\mathbb{W}_\mathrm{in}$ are selected from $\{-1,1\}$ (\textbf{C}) and $\{-2,2\}$ (\textbf{D}) respectively. The white dash-dotted curves denote the contour lines of $v_0\delta t/R_0$. The yellow dashed curves denote the transition points of $U_\mathrm{eff}(\theta)$ from single well to double well form. An instrumental feedback latency $\delta t_F$ = 0.125 s is considered, thereby the transition curves do not coincide with the $v_0\delta t/R_0$ contour lines. 
    For more details see Sec.~S1--S2 of Supplementary Information. The subplot on the right illustrates the corresponding $U_\mathrm{eff}(\theta)$, also $\theta$ perturbations and relaxations. 
    \textbf{E} Comparison of NRMSE results from panel (\textbf{C}) and (\textbf{D}) as function of $v_0\delta t/R_0$ with $v_0/R_0 = 1.01$. The dashed line denotes the transition point of $U_\mathrm{eff}$. }
    \label{fig:5}
\end{figure}

\paragraph*{\bf Impact of noise}
These results suggests, that there might be an optimal relation between the used number of historical states and the error of the RC to stabilize the prediction and to reduce the impact of the inherent noise due to Brownian motion. To investigate this interelation we refer to deterministic and stochastic simulations. Fig.~\ref{fig:5}A,~B display the RC performance as measured by the NRMSE of the predictions as functions of $P_\mathrm{in}$ and $P_\mathrm{out}$. Without noise (Fig.~\ref{fig:5}A), the results indicate a larger error with low $P_\mathrm{in}$ or $P_\mathrm{out}$ and poor performances for $P_\mathrm{in}<$ 2 and $P_\mathrm{out}<$ 50. The performance is improving for increasing $P_\mathrm{in/out}$ and with an approximate linear correlation to obtain optimal results Fig.~\ref{fig:5}A. Best results for the deterministic system are obtained with $n_\mathrm{hist}=P_\mathrm{out}/P_\mathrm{in}-1=60$. 

For the noisy active particles nodes (Fig.~\ref{fig:5}B), the RC outputs are unstable for $P_\mathrm{out}<100$. The free-running prediction has a high probability to yield fast diverging outputs to yield a large NRMSE (bottom subplot). 
With $P_\mathrm{out}\geq 100$ a stable output is achieved. A trend towards lower NRMSE with higher $n_\mathrm{hist}$ can be observed in agreement with our experimental results (Fig.~\ref{fig:4}F). The results again suggest a linear relation between $P_{out}$ and $P_{in} $ to obtain a minimum error. Thus, both experiments and simulations confirm that using historical states $n_\mathrm{hist}$ for the prediction of our RC improves the stability and the quality of the prediction even under extremely noisy conditions. 

\paragraph*{\bf Impact of the dynamical properties of the nodes}
Our active particle recurrent node also provides the opportunity to tune the dynamical response of the physical node across the transition of the pitchfork bifurcation with either particle speed $v_0$ or delay $\delta t$. The tuning varies the memory of each node as indicated in Fig. \ref{fig:2}D by the relaxation time $\tau_{\theta}$ and thereby also the coupling of the virtual nodes as stated by Eq. \ref{eq:theta evolution}. 

The impact of this change on the performance of the RC is depicted in Fig.~\ref{fig:5}C--E  with the NRMSE for the MGS predictions as function of $v_0/R_0$ and $\delta t$ resulting from deterministic simulations. As discussed above, $\theta_0$ determines the asymptotic behavior after an impulsive perturbation of $\theta$. The optimum around $\theta_0\approx 0.2$ and the correlation with $\theta_0={\rm const.}$ indicate that the dynamical properties of the physical node, in particular the fading memory as characterized by the relaxation time $\tau_{\theta}$ (Fig.~\ref{fig:2}B,~D) are indeed a key factor of the RC performance for this task. 
The optimal performance of the deterministic system is found below the transition point (yellow line) where the approximate effective potential is largely determined by the parabolic part (Fig.~\ref{fig:5}C, right)~\cite{Wang2023}. While the relaxation time is largest at the transition point and an input $u$ perisists for the longest time, the NRMSE is found to be largest considerbly beyond the transition point as indicated by the yellow dashed line in (Fig. \ref{fig:5} C). In this region, the effective potential is a double well potential allowing for an inconsistent response, i.e. an input perturbation may result in a relaxation into either of the minima. If the barrier of the double well potential increases further the NRMSE appears to decrease again (Fig.~\ref{fig:5}E) presumably due to the fact that active particle node is largely residing in one of the two potential wells only with a short relaxation time.

\section{Discussion}

We have demonstrated above the realization of a physical reservoir computer in an experiment using self-propelled active microparticles. A retarded propulsion towards an immobilized target particle creates a self-organized non-linear dynamical system that is, despite the strong Brownian noise, capable of predicting chaotic time series such as Mackey-Glass and Lorenz series when used as a physical node in a reservoir computer. The key element of this physical recurrent node is a time-delay realizing a retarded interaction~\cite{Mijalkov.2016,Wang2023,Chen.2023} that creates the fading memory as a basic requirement for reservoir computing~\cite{Jaeger2001}. It also provides the coupling of the active particle dynamics to its past allowing to implement virtual nodes living on a single physical active particle system via time-multiplexing~\cite{Appeltant2011,Appeltant2014}. The information processing that is provided by such a single node may therefore extend the rare simulation work on reservoir computing with active particle swarms with interparticle coupling~\cite{Lymburn2021}. Additionally, the nonlinearity that is required for computations is an intrinsic physical property of our active particle system and requires no extra treatment of the output signal ~\cite{Cucchi2022}. Future work could introduce direct physical coupling between the isolated physical nodes in our configuration, e.g., through hydrodynamic or other interactions, to obtain more complex dynamical networks of interacting synthetic active particles.

The dynamics of the physical recurrent node that is the basic unit of our reservoir computer is controlled by a parameter containing the product of activity (active particle speed) and time-delay and can be understood with a simple Landau-like self-induced quartic effective potential, which exhibits a bifurcation to a double well potential~\cite{Wang2023}. The system thereby allows to address the relation of reservoir performance and node relaxation time, were we found a clear indication for an optimal performance for small delays below the bifurcation point in deterministic simulations. While the fading memory becomes extremely long at the bifurcation and one might expect the worst performance of the system, it is observed that a double well potential with a small barrier leads to inconsistencies in the relaxation dynamics that are more severe even in the deterministic system. Interestingly, such double well potentials have recently been discussed as non-linear stochastic-resonance-based activation functions in an attempt to provide better stability of Echo-State-Networks against noise~\cite{Guo2018,Liao2021}. 

Noise is an inherent property of our information processing units, as Brownian motion causes strong fluctuations of the node state. Such noises are inevitable at the smallest scales also in the context of biological information processing~\cite{Tsimring2014}, for instance in neurons~\cite{Faisal2008}, both with positive and negative effect~\cite{Stein2005,Guo2018}. In physical RC approaches, noise is commonly a major limiting factor for the performances~\cite{Dambre2012,Soriano2013,Antonik2017b} although subtle noises are reported to be beneficial as well~\cite{Jaeger2001,Jaeger2004,Sussillo2009,Paquot2012,Estebanez2019}). Yet, general strategies for the noise suppression are unclear except increasing the reservoir size~\cite{Alata2020}, which is normally costly for physical RCs. While the performance of our reservoir computer for the chaotic system prediction is highly degraded by the noise due to the sensitivity of chaotic systems, we have introduced the output generation including historical node states providing remarkable stability and noise reduction evenunder low signal to noise ratios (see Sec.~S7.1 of Supplementary Information). This architecture is not increasing the dimensionality of the reservoir, but gives different weight to contributions of reservoir states from the past for the current output and could be potentially useful in future reservoir computing studies.   

In summary, simple retarded interactions in synthetic active microparticle systems can give rise to non-linear self-driven dynamics that form a basis for information processing with active matter. Our reservoir computer highlights this connection between information processing, machine learning and active matter on the microscale and paves the way for new studies on noise in reservoir computing. While we so far referred to isolated active recurrent units, we envision that the high level of control of synthetic active matter design will yield new emergent physical collective states that may leverage the field of active synthetic dynamical systems for information processing.

\section*{Methods}
\label{sec:methods}
\subsection{Sample preparation} 
The sample used in experiments contains two kinds of micro particles: polystyrene (PS) particles (microParticles GmbH) of $3~\mum$  diameter and the melamine formaldehyde (MF) (microParticles GmbH) particles of $2.19~\mum$ diameter, suspended in a water solution (Fig.~\ref{fig:1}B). Gold (Au) nano-particles of around $8~\si{\nano\metre}$ diameter uniformly distributed on the surface of the MF particle, cover about 10\% of the total surface area of the latter. Two glass coverslips ($20\,\times\,20~\si{\milli\metre}^2$ and $24\,\times\,24~\si{\milli\metre}^2$) confine a $3~\mum$ thick sample layer in between. Due to the surface tension of water, the PS particles are compressed and immobilized on the coverslips serving as spacers to define the sample thickness. The PS and MF-Au particles are separately added into two 2\% Pluronic F-127 solutions. After 30 minutes, the Pluronic concentration of both solutions is decreased to 0.02\% by diluting twice. Each dilution is followed by an centrifugation then a removal of part of the solution to keep the particle concentration. 
$0.3~\si{\micro\litre}$ sample of PS particles is pipetted on one of the coverslips, then $0.3~\si{\micro\litre}$ sample of MF-Au particles is pipetted in the droplet of PS particles. The sample is then covered carefully with a second coverslip. The edges of the sample are sealed by polydimethylsiloxane (PDMS) to prevent leakage and evaporation, also to relieve the liquid flow inside the sample. The experiment starts about one hour after sample preparation to wait until residual liquid flows in the sample ceased.

\subsection{Experimental setup} 

The experimental setup is illustrated in Fig.~\ref{fig:1}A. The micro MF-Au particles are heated by a focused, continuous-wave laser with a wavelength of $532~\si{\nano\metre}$. The light from the laser module (CNI, MGL-H-532-1W) is expanded by two tube lenses ($35, 150~\si{\milli\metre}$ focal lengths) in the beam size, then guided by mirrors to a high-speed reflective Spacial Light Modulator (SLM, Meadowlark Optics, HSP512-532), which modulates the phase of the reflected laser. The reflected laser is then guided through two lenses ($500, 300~\si{\milli\metre}$ focal lengths) to an inverted microscope (Olympus, IX73). A small opaque dot on a glass window located at the focal point of the $500~\si{\milli\metre}$ lens, serves as a mask to block the unmodulated laser reflected by the SLM top surface. The laser in the microscope is reflected by a dichroic beam splitter (Omega Optical, 560DRLP), then focused by an objective lens (100x, Olympus, UPlanFL N x100/1.30, Oil, Iris, NA. 0.6--1.3) on the sample plane with a beam width at half maximum about $0.6~\mum$. 

The sample is illuminated by white light from a LED lamp (Thorlabs, SOLIS-3C) through an oil-immersion dark-field condenser (Olympus, U-DCW, NA 1.2--1.4).
The image of the sample is projected by the objective lens and a tube lens ($180~\si{\milli\metre}$ focal length) inside the microscopy stand as well as two additional lenses ($100, 150~\si{\milli\metre}$ focal lengths) outside the microscopy stand to a camera (Hamamatsu digital sCMOS, C11440-22CU). The numerical aperture (NA) of the objective is set to a value below the minimal NA of the dark-field condenser. Two filters (EKSMA Optics 246-2506-532, Thorlabs FESH0800) in front of the camera block the back reflections of the laser from the microscope. A desktop PC (Intel(R) Core™ i7-7700K CPU @4$\times$4.20 GHz, NVIDIA GeForce GTX 1050Ti) with a LabVIEW program (v.~2019) analyses the images, records data, and manipulates the active particles by controlling the laser through the phase pattern on the SLM. More details are given in Sec.~S1 of Supplementary Information.

\section*{Acknowledgement}
This work is supported by Center for Scalable Data Analytics and Artificial Intelligence (Scads.AI) Dresden/Leipzig.

\section*{Competing interests}
The authors have no competing interests.

\section*{Author Contributions}

\section*{Data availability}
All data in support of this work is available in the manuscript or the supplementary materials. Further data and materials are available from the corresponding author upon request.


\end{document}